\documentclass[aps,prl,reprint,groupedaddress,twocolumn]{revtex4-2}

\usepackage{graphicx}
\usepackage[colorlinks,linkcolor=blue,anchorcolor=black,citecolor=blue]{hyperref}
\usepackage{dcolumn}
\usepackage{bm}
\usepackage{autobreak}
\usepackage{url}
\usepackage{booktabs} 
\usepackage{amsmath}
\usepackage{threeparttable}
\usepackage{multirow}

\newcommand{\insitu}{\textit{in situ }}

\begin{document}


\title{\textit{In situ} Qubit Frequency Tuning Circuit for Scalable Superconducting Quantum Computing: 
Scheme and Experiment}


\author{Lei Jiang$^{1,2}$}
\thanks{These authors contributed equally to this work.}
\author{Yu Xu$^{2,3}$}
\thanks{These authors contributed equally to this work.}
\author{Shaowei Li$^{2,3}$}
\thanks{These authors contributed equally to this work.}
\author{Zhiguang Yan$^{1,2}$}
\thanks{Present address: RIKEN Center for Quantum Computing (RQC), Wako, Saitama 351-0198, Japan}
\author{Ming Gong$^{1,2,3}$}
\author{Tao Rong$^{1,2}$}
\author{Chenyin Sun$^{1,2}$}
\author{Tianzuo Sun$^{1,2}$}
\author{Tao Jiang$^{1,2}$}
\author{Hui Deng$^{1,2,3}$}
\author{Chen Zha$^{2,3}$}
\author{Jin Lin$^{2,3}$}
\author{Fusheng Chen$^{2,3}$}
\author{Qingling Zhu$^{2,3}$}
\author{Yangsen Ye$^{1,2}$}
\author{Hao Rong$^{1,2}$}
\author{Kai Yan$^{2,3}$}
\author{Sirui Cao$^{1,2}$}
\author{Yuan Li$^{1,2}$}
\author{Shaojun Guo$^{1,2}$}
\author{Haoran Qian$^{1,2}$}
\author{Yisen Hu$^{1,2}$}
\author{Yulin Wu$^{1,2}$}
\author{Yuhuai Li$^{1,2,3}$}
\author{Gang Wu$^{1,2,3,4}$}
\author{Xueshen Wang$^{3}$}
\author{Shijian Wang$^{3}$}
\author{Wenhui Cao$^{3}$}
\author{Yeru Wang$^{5}$}
\author{Jinjin Li$^{3}$}
\email{jinjinli@nim.ac.cn}
\author{Cheng-Zhi Peng$^{1,2,3}$}
\author{Xiaobo Zhu$^{1,2,3,5}$}
\email{xbzhu16@ustc.edu.cn}
\author{Jian-Wei Pan$^{1,2,3}$}
\email{pan@ustc.edu.cn}
\affiliation{$^1$Hefei National Research Center for Physical Sciences at the Microscale and School of Physical Sciences, University of Science and Technology of China, Hefei 230026, China}
\affiliation{$^2$Shanghai Research Center for Quantum Science and CAS Center for Excellence in Quantum Information and Quantum Physics, University of Science and Technology of China, Shanghai 201315, China}
\affiliation{$^3$Hefei National Laboratory, University of Science and Technology of China, Hefei 230088, China}
\affiliation{$^4$University of Science and Technology of China, Shanghai Research Institute, Shanghai 201315, China}
\affiliation{$^5$Jinan Institute of Quantum Technology, Jinan 250101, China}


\date{\today}

\begin{abstract}

Frequency tunable qubit plays a significant role for scalable superconducting quantum processors. The state-of-the-art room-temperature electronics for tuning qubit frequency suffers from unscalable limit, such as heating problem, 
linear growth of control cables, etc. Here we propose a scalable scheme to tune the qubit frequency by using \insitu superconducting circuit, which is based on radio frequency superconducting 
quantum interference device (rf-SQUID). We demonstrate both theoretically and experimentally that the qubit frequency could be modulated by inputting several single pulses into rf-SQUID. 
Compared with the traditional scheme, our scheme not only solves the heating problem, but also provides the potential to exponentially reduce the number of cables inside the dilute refrigerator and the room-temperature electronics resource for tuning qubit frequency, which is achieved by a time-division-multiplex (TDM) scheme combining rf-SQUID with switch arrays. With such TDM scheme, the number of cables could be reduced from the usual $\sim 3n$ to $\sim \log_2{(3n)} + 1$ for two-dimensional quantum processors comprising $n$ qubits and $\sim 2n$ couplers. 
Our work paves the way for large-scale control of superconducting quantum processor.

\end{abstract}


\maketitle

\section{Introduction}

Recent years quantum advantage has been realized on superconducting quantum computing system built upon 
frequency tunable superconducting qubit\cite{google_2019_supremacy,ustc_2021_advantage,ustc_2022_advantage,
google_RCS_2023}, due to its remarkable flexibility of fulfilling various quantum gates with high fidelity 
\cite{Google2014_threshold,Oliver_2018tunablecoupling,UCSB_2019diabaticgates,Google_2021Reset}, 
controlling the coupling strength between qubits when serving as couplers \cite{UCSB2011_tunablecoupler,
Oliver_2018tunablecoupling}, etc. 
Currently, tuning the transition frequency of frequency tunable qubit i.e., Z-control is achieved by inputting a 
constant direct current (DC) signal or a long-time pulsed square wave microwave signal from room-temperature electronics 
into the low temperature qubit chip. However, this scheme suffers from the heating problem caused by consumed energy on 
the attenuators along the coaxial cables within the dilute refrigerator\cite{ustc_2021_advantage,google_2019_supremacy}. 
And the number of Z-control line
increases linearly with number of qubits and couplers, challenging the limited inner space of the dilute refrigerator 
when scaling up to thousands of qubits. These factors contribute to the bottleneck of the scalability of superconducting quantum 
processors.

Here we propose a scalable scheme to tune the qubit frequency by \insitu superconducting loop based on 
radio frequency superconducting quantum interference device (rf-SQUID). Today rf-SQUID is prevalently used as magnetic 
flux and field sensor to detect the external magnetic field \cite{squid_advance_1980,squid_handbook_2004}. 
And applying rf-SQUID to generate magnetic flux has also been investigated since 1970s \cite{smith_1975,
smith_1987,robert_1978}. There are also many works applying rf-SQUID to superconducting quantum computing e.g., 
rf-SQUID directly acting as qubit \cite{rfsquid_qubit_2000,rfsquid_qubit_2005_Science,D_wave_2011_nature}, 
using rf-SQUID to detect quantum states of qubits \cite{rfsquid_readout_2004,rfsquid_readout_2004_prl},
rf-SQUID acting as tunable coupler \cite{rfsquid_coupler_2010_prl,Oliver_2019_apl}, etc. 
However, applying rf-SQUID to control qubit frequency in a scalable way has rarely been reported. 

In this paper, we propose a scalable scheme to control qubit frequency by utilizing rf-SQUID as a magnetic flux generator. 
In our scheme an rf-SQUID is designed around a qubit to provide qubit with magnetic flux via mutual 
inductance. We will demonstrate both theoretically and experimentally that the state of rf-SQUID i.e., superconducting 
current inside rf-SQUID could be modulated by a pulsed signal. The state of rf-SQUID will stay the same after modulation 
indefinitely due to the thermal excitation rate approaching zero \cite{SM}, thus providing qubits
with constant magnetic flux to keep qubit frequency stable. In experiment, we also compare the stability of qubit frequency
obtained in classical room-temperature electronics scheme and the rf-SQUID scheme. The relative standard deviation 
(RSD) of magnetic flux fluctuation provided for qubit under rf-SQUID scheme is 24.5 parts per million (PPM),
outperforming 55.0 PPM observed under room-temperature electronics scheme (driven by 16-bit resolution digital-to-analog 
converter\cite{ustc_2021_advantage}). Compared with the classical
Z-control scheme based on room-temperature electronics, our scheme not only requires several single, short-duration pulsed signals 
input to tune qubit frequency, which could overcome the heating problem rising from the continuous signal input, but also 
could further reduce the room-temperature electronics resource for Z-control and the number of Z-control cables in 
dilute refrigerator from usual $\sim 3n$ to $\sim \log_2(3n)+1$ for quantum processor with $n$ qubits and $\sim 2n$ 
couplers, so as to move forward to low-power and scalable control of large-scale superconducting quantum processors.

\section{the model and scheme}
\begin{figure*}[htbp]
	\includegraphics[keepaspectratio,width=17.5cm]{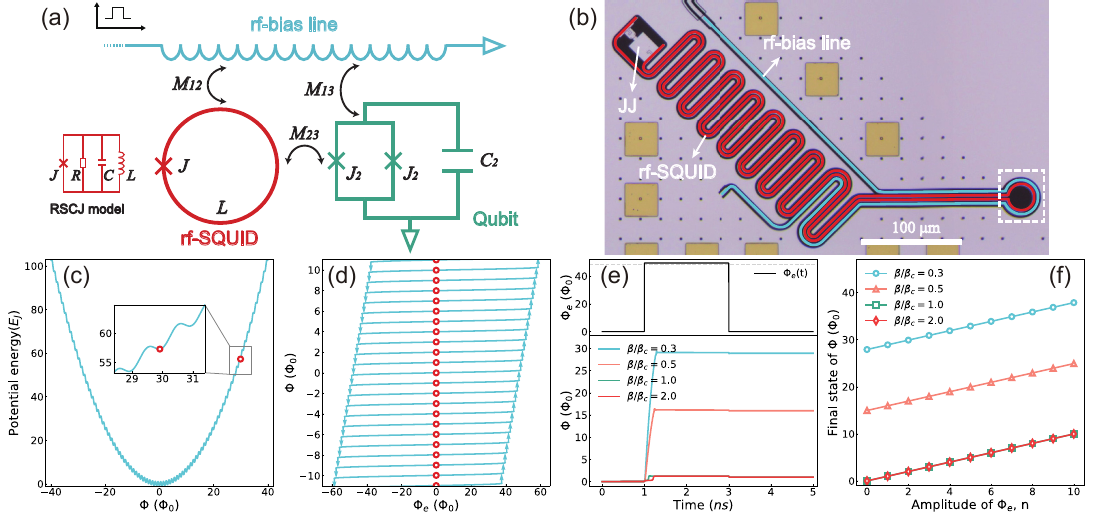}
	\caption{
    Tuning qubit frequency by rf-SQUID. 
		(a) Schematic circuit diagram of tuning the transition frequency of transmon by an rf-SQUID.
		(b) Optical micrograph of the rf-SQUID on the bottom chip of our flip-chip sample. The red line represents the inductance loop of 
		rf-SQUID which is connected by a JJ and the blue line outlines rf-bias line. The white dashed square 
		represents the area where the rf-SQUID and the rf-bias line interact with qubit (not shown) via spatial inductive coupling.
		(c) A numerical example of the potential energy curve of the rf-SQUID when external magnetic flux $\Phi_e=0$ 
		for $I_c$=100 $\mu$A, $L$=1 nH, and $\beta_e=303.9$. The horizontal axis is the total magnetic flux $\Phi$ of rf-SQUID
		and the vertical axis represents the potential energy $U$ in the unit of Josephson energy $E_J=I_c \Phi_0/2\pi$. 
		The inset figure shows the detail of the local tilted washboard potential well located at $\Phi\approx$30 $\Phi_0$.
		(d) A numerical example of hysteresis curve between external magnetic flux $\Phi_e$ and total magnetic flux $\Phi$ of 
		rf-SQUID with the same parameters settings in (c). The red circles indicate the position of local metastable 
		states for $\Phi_e$=0 which are consistent with those shown in (c).
		(e) A numerical example of modulating the state of rf-SQUID by inputting a single pulsed signal for different damping 
		condition. The parameters in simulation are chosen $I_c$=100 $\mu$A, $L$=1 nH, $C$=2 pF and $R$=3.6 $\Omega$, 2.2 $\Omega$, 
		1.1 $\Omega$, 0.5 $\Omega$ for $\beta/\beta_c$= 0.3, 0.5, 1.0, 2.0, respectively. The initial state of rf-SQUID is 
		set to $\Phi=0$. The amplitude of the pulse is chosen as $\Phi_e=(\beta_e/2\pi+1)\Phi_0=49.4 \Phi_0$ and the
		critical value of $\Phi_e$ when $\Phi=0$ is $\Phi_{e,c}=(\beta_e/2\pi + 1/4)\Phi_0=48.6\Phi_0$, which is shown 
		by the light grey dashed line in the upper subfigure.
		(f) A numerical example of the linear relationship of magnetic flux transition of rf-SQUID for different damping conditions with 
		the same simulation parameters settings in (e). The horizontal axis represents the amplitude of the pulse in simulation
		$\Phi_e=[\beta_e/(2\pi)+n]\Phi_0$ with $n=0,1,...,10$. The result of $\beta/\beta_c$=1.0 coincides with that of
		$\beta/\beta_c$=2.0. All results share a same slope $k=1$.
	}    
	\label{fig:SQUID principle}
\end{figure*}
Fig. \ref{fig:SQUID principle}(a) shows the circuit diagram of our scheme. There is an rf-SQUID (red) 
locating around qubit (green) which interacts with qubit through mutual inductance $M_{23}$. To modulate the 
state of rf-SQUID, we design a transmission line named rf-bias line (blue) around rf-SQUID which not only interacts 
with rf-SQUID by mutual inductance $M_{12}$ but also directly interacts with qubit through mutual inductance $M_{13}$. 
In this way, rf-bias line also works as the microwave drive line (XY-line) for qubit. 
A single square wave pulse is input into rf-bias line to modulate the state of rf-SQUID 
i.e., the superconducting current inside the loop of rf-SQUID. After modulation the superconducting current inside 
rf-SQUID provides qubit with magnetic flux through $M_{23}$, which tunes the qubit frequency.

In this scheme, we focus on the rf-SQUID with hysteresis parameter (screening parameter) 
$\beta_e=2\pi \frac{I_c L}{\Phi_0}>1$, where $I_c$ is the critical Josephson current of the Josephson junction (JJ) 
in rf-SQUID, $L$ is the self-inductance of the superconducting loop and $\Phi_0=h/2e\approx 2.07 $ $mV\cdot ps$ is the magnetic flux quantum. 
When $\beta_e>1$, the potential energy of rf-SQUID will exhibit many local potential wells (metastable states) as shown in Fig \ref{fig:SQUID principle}(c) and the local potential wells $\Phi_m$ has the approximation $\Phi_m \approx n \Phi_0$ with 
$n=0,\pm 1,\pm 2,...,\pm \lfloor \beta_e/2\pi +1/4\rfloor$ (see Eq. \ref{equ:critical value of Phi and Phie}). Moreover, the relationship between the external magnetic 
flux $\Phi_e$ and the total magnetic flux $\Phi$ of rf-SQUID will exhibit hysteresis as shown in Fig. \ref{fig:SQUID principle}(d), which is indicated by Eq.\ref{equ:static equation}. In this case, the rf-SQUID is 
referred to as hysteretic \cite{physics_of_JJ_1982,dynamics_of_JJ,squid_handbook_2004}. 
Since $\Phi=\Phi_e + IL$, the total magnetic flux 
$\Phi$ of rf-SQUID i.e., the state of rf-SQUID determines the superconducting current $I$ inside rf-SQUID for $\Phi_e=0$. 
\begin{equation}
	\frac{\Phi_e}{\Phi_0} = \frac{\Phi}{\Phi_0} + \frac{\beta_e}{2\pi}\sin\Big{(}2\pi\frac{\Phi}{\Phi_0}\Big{)}
	\label{equ:static equation}
\end{equation}

\begin{equation}
	\begin{split}
		&\Phi_{e,c} =\big{[}n \pm (\beta_e/2\pi +1/4)\big{]}\Phi_0\\
		&\Phi_m \approx n\Phi_0\\
		&n=0,\pm 1,\pm 2,...,\pm \lfloor \beta_e/2\pi +1/4\rfloor
	\end{split}
	\label{equ:critical value of Phi and Phie}
\end{equation}

The state of rf-SQUID could be modulated by a pulsed signal. The dynamical differential equation rises from the 
resistive-capacitive shunted Josephson junction (RSCJ) model of rf-SQUID (see Fig \ref{fig:SQUID principle}(a)) 
\cite{physics_of_JJ_1982,dynamics_of_JJ,squid_handbook_2004}
\begin{align}
	\begin{autobreak}
		\frac{d^2(\Phi/\Phi_0)}{dt^2} = 
		-\frac{\beta}{\sqrt{LC}} \frac{d(\Phi/\Phi_0)}{dt} 
		- \frac{\beta_e}{2\pi} \frac{1}{LC}\sin\Big{(}2\pi \frac{\Phi}{\Phi_0}\Big{)} 
		+ \frac{1}{LC}\Big{(}\frac{\Phi_e(t) - \Phi}{\Phi_0}\Big{)}
	\end{autobreak}
	\label{equ:dynamical differential equation}
\end{align} 
where $C$ is the shunted capacitance of the JJ, $R$ is the shunted resistance of the JJ, $L$ is the self-inductance 
of the loop and $\beta=\frac{1}{R}\sqrt{\frac{L}{C}}$ is the quality factor of rf-SQUID. With the help of Eq.
\ref{equ:dynamical differential equation}, it's straightforward to deduce the dynamical response of rf-SQUID to a pulsed external magnetic 
flux $\Phi_e(t)$ when $\beta/\beta_c > 1$ (overdamped condition), $\beta/\beta_c = 1$ (critical damped condition)
or $\beta/\beta_c < 1$ (underdamped condition) where $\beta_c=2.97\sqrt{\beta_e/2\pi}$, as shown in 
Fig. \ref{fig:SQUID principle}(e) \cite{physics_of_JJ_1982,dynamics_of_JJ,smith_1975}. These dynamical results could be 
illustrated by a physical picture that a particle with mass $C$ slips down a tilted washboard potential once the local 
potential barrier vanishes when the input magnetic flux signal exceeds the threshold value $\Phi_{e,c}$ (see Eq. 
\ref{equ:critical value of Phi and Phie} deriving from Eq. \ref{equ:static equation}). And the final stop position 
is related with the damping condition $\beta/\beta_c$. Additionally, the final state of rf-SQUID varies
linearly with the amplitude of input magnetic flux once the amplitude exceeds the thresholds regardless of the damping 
conditions, which is shown in Fig. \ref{fig:SQUID principle}(f). This linear magnetic flux transition relationship is 
crucial to modulate rf-SQUID to reach any desired local potential well. All these dynamical properties inspire us the scheme
of modulating rf-SQUID by a pulsed signal.

For $\beta/\beta_c \geq 1$(overdamped or critical damped condition), the amplitude of the input pulse should be chosen near
the threshold value indicated by the target position. To be specific, let's suppose the initial state of rf-SQUID is 
$\Phi_i \approx n_i \Phi_0, n_i >0$ and the target final state is $\Phi_f \approx n_f \Phi_0$ with $ n_f > n_i$. 
In this case, the pulse amplitude must satisfy $\Phi_e/\Phi_0\in \big{[}n_f-1 + (\beta_e/2\pi +1/4),n_f + (\beta_e/2\pi +1/4)\big{)}$
and vice versa for other $n_i,n_f<0$ cases. Only a single pulse input is needed in principle.

For $\beta/\beta_c <1$(underdamped condition), it's usually hard to modulate rf-SQUID to a specific position
by inputting a single pulse due to the multiple magnetic flux transition phenomenon (see Fig. \ref{fig:SQUID principle}(e))
\cite{smith_1975,physics_of_JJ_1982}. However, it's still able to modulate rf-SQUID to a target position by inputting 
several single pulses. For example, let's suppose that in this case, the basic magnetic flux transition $\Delta \Phi$
satisfies $\Delta \Phi>0$. Firstly, inputting a single pulse with negative amplitude $\Phi_{e,1}/\Phi_0\in(n_i-1 - (\beta_e/2\pi +1/4),n_i - (\beta_e/2\pi +1/4)]$,
rf-SQUID will evolve into a temporal position $\Phi_r\approx n_r \Phi_0$ which satisfies $|\Phi_r-\Phi_i|=\Delta \Phi$.
Secondly, another input single pulse with positive amplitude $\Phi_{e,2}/\Phi_0\in\big{[}n_r+n_f-n_i + (\beta_e/2\pi +1/4),n_r+n_f-n_i+1+ (\beta_e/2\pi +1/4)\big{)}$
(deduced from the linear magnetic flux transition relationship) will modulate rf-SQUID to the target state $\Phi_f$. There are similar strategies for $n_f<n_i$ cases.

\section{Design and Experiment}
Fig. \ref{fig:SQUID principle}(b) shows an optical micrograph of the rf-SQUID of our sample, where the red line represents 
the inductance loop of rf-SQUID which is connected by a JJ and the blue line outlines the rf-bias line. The chip has a flip-chip structure which is fabricated by the micro-nano fabrication technology developed in our \textit{Zuchongzhi} series quantum processor \cite{ustc_2021_advantage}. 
\begin{table}[htbp]
	\centering
	\small
	\begin{threeparttable}
	\caption{
		Parameters of rf-SQUID
		\label{tab:design parameters of rf-SQUID}}
	\begin{tabular}{lll}
		\toprule 
		Parameters & $\qquad$ & Value\\ 
		\midrule
		Josephson current $I_c$ & $\qquad$ & 320 $\mu$A \\
		Self-inductance $L$ & $\qquad$ & 1.18 nH \\
		Mutual inductance $M_{12}$ & $\qquad$ &  70 pH\\
		Mutual inductance $M_{13}$ & $\qquad$ &  2.90 pH \\
		Mutual inductance $M_{23}$ & $\qquad$ &  4.07 pH \\
 		\bottomrule 
	\end{tabular} 
	\small
	\end{threeparttable}
\end{table}

\begin{figure*}[htbp]
	\includegraphics[keepaspectratio,width=17.5cm]{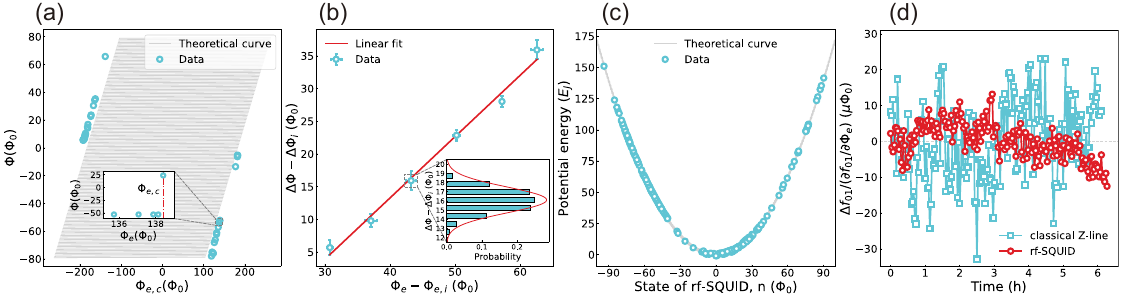}
	\caption{
 Experimental results of modulating rf-SQUID states by single square wave pulses.
		(a) The hysteresis result between critical external magnetic flux $\Phi_{e,c}$ and states $\Phi$ of rf-SQUID. The 
		experimental data is shown in blue circles and the theoretical curve is shown in light grey lines. The inset
		figure shows an example that rf-SQUID does not change from the initial position -53.2 $\Phi_0$ to the
		final position 23.5 $\Phi_0$ until the amplitude of the pulse reaches its critical value $\Phi_{e,c}=138.4\Phi_0$.
		(b) The result of magnetic flux transition experiment. 
		We repeatedly prepare the initial state $\Phi_i=(-57.5\pm 1)\Phi_0$ for rf-SQUID, whose critical amplitude 
		of the pulse is $\Phi_{e,i}=(94.9\pm 1)\Phi_0$. We vary the amplitude $\Phi_e$ and collect 
		the final state $\Phi_f$ of rf-SQUID.
		The horizontal axis represents the difference between the amplitude $\Phi_e$ of input pulse and the critical amplitude 
		$\Phi_{e,i}=(94.9\pm 1)\Phi_0$ of the pulse of the initial state. The vertical axis represents the difference
		$\Delta \Phi - \Delta \Phi_i$ of the magnetic flux transition, where $\Delta \Phi=\Phi_f-\Phi_i$,
		$\Delta \Phi_{i} = \Phi_{f,i}-\Phi_i=76.7\Phi_0$ and $\Phi_{f,i}$ represents the final state of rf-SQUID when 
		inputting pulse with critical amplitude $\Phi_{e,i}$.
		Each data (blue circle) is sampled for at least 4 times. The linear fit result
		$\Delta\Phi - \Delta \Phi_i=k(\Phi_e-\Phi_{e,i}) +b$ with $k=0.94(0.05)$ and $b=-24.23(2.17)$ demonstrates that the 
		data aligns well with theoretical linear trend. The inset figure shows the detailed distribution of the magnetic flux 
		transition at the 3rd data point. The result contains 626 data and is fit by Gaussian function
		$p=1/(\sqrt{2\pi}\sigma) e^{-(x-x_0)/2\sigma^2}$ with $x_0=16.1(0.1)$ and $\sigma=1.4(0.1)$.
		(c) The statistics of the local potential wells of rf-SQUID by varying the amplitude of single square wave pulse 
		in the experiment. The horizontal axis represents the state $n$ of rf-SQUID, and the vertical axis represents the potential energy
		in the unit of Josephson energy $E_J$. The result contains total 171 data (bule circles).
		(d) Comparison of the stability results of qubit frequency $f_{01}$ collected under rf-SQUID scheme and classical Z-line
		scheme, respectively. The red circles represent the data collected under rf-SQUID scheme with $f_{01}=4.5904$ GHz 
		whose maximum $f_{01}$ is 5.1387 GHz. The blue squares are the data collected under the classical Z-line scheme with 
		$f_{01}=4.4727$ GHz whose maximum $f_{01}$ is 5.0069 GHz. 
		Each data point is sampled for every 2 minutes. The y-axis represents the relative qubit frequency $\Delta f_{01}$ difference from the mean $f_{01}$, divided by the 
		derivative of $f_{01}$ versus external magnetic flux $\Phi_e$ for qubit. The peak-to-peak (P2P) value and standard 
		deviation (STD) of $f_{01}$ for rf-SQUID case are $\text{P2P}=25.7$ $\mu \Phi_0$ and $\text{STD}=4.9$ $\mu \Phi_0$, while for the
		classical Z-line case, these results are $\text{P2P}=56.0$ $\mu \Phi_0$ and $\text{STD}=11.0$ $\mu \Phi_0$. 
	}    
	\label{fig:SQUID results}
\end{figure*}
In the experiment, we firstly use the rf-bias line as XY-drive line and Z-control line to calibrate qubits after the 
chip installed into a dilute refrigerator reaching temperature below 20 mK. 
After calibrating basic qubit properties, we input single square wave voltage signal from room-temperature electronics into the rf-bias line to modulate the state of rf-SQUID.
The pulsed signal 
typically has an amplitude of 7 V and a duration of 100 ns, which goes through total 34 dB attenuators along coaxial cables in 
the dilute refrigerator. After modulation, no more signal to tune qubit frequency is input and we deduce the state information of rf-SQUID from qubit frequency calibrated by Ramsey experiment.

Table \ref{tab:design parameters of rf-SQUID} lists the  
parameters of the rf-SQUID in the experiment. These parameters enable total $2\times \lfloor \beta_e/2\pi+1/4 \rfloor-1=363$ 
local potential wells and maximum magnetic flux 0.63 $\Phi_0$ provided for qubit. 
The difference in the magnetic flux between the adjacent wells of rf-SQUID approximates 1 $\Phi_0$, 
thus the magnetic 
flux step providing for qubit is 0.0034 $\Phi_0$ under these parameters, which is sufficient for 
traversing qubit frequencies. Moreover, combining the data from current-voltage (I-V) curve of 
the JJ in rf-SQUID, the damping condition of rf-SQUID satisfies $\beta=12.1<\beta_c=40.1$, which indicates the 
underdamped condition of rf-SQUID \cite{SM}.

Since the rf-SQUID is underdamped, we scan the hysteresis curve by alternately modulating the state $\Phi$ of rf-SQUID 
back and forth between positive and negative positions, which is achieved by inputting corresponding single pulses with negative 
and positive amplitudes. Fig. \ref{fig:SQUID results}(a) shows the experimental result. 
Here the vertical axis is the state $\Phi$ of rf-SQUID deduced from qubit frequency. 
The horizontal axis represents the critical amplitude $\Phi_{e,c}$ of the pulse which begins to modulate rf-SQUID. 
And the inset figure exhibits an example 
that rf-SQUID does not change from the initial position -53.2 $\Phi_0$ to the final position 23.5 $\Phi_0$ until the amplitude of the pulse 
reaches its critical value 138.4 $\Phi_0$.

We then perform the linear magnetic flux transition experiment with the initial state of rf-SQUID repeatedly prepared to the target 
position $\Phi_i=(-57.5\pm 1)\Phi_0$. After preparation, we vary the amplitudes of single square wave voltage pulses 
to investigate the magnetic flux transition relation.
In the experiment, the positive critical amplitude of $\Phi_i$ is $\Phi_{e,i}=(94.9\pm 1)\Phi_0$ which is 
different from the theoretical prediction $\Phi_{e,i,predict}=(125.4\pm 1)\Phi_0$. We attribute this difference 
to the environmental magnetic flux around rf-SQUID when we perform the experiment \cite{SM}.
Fig. \ref{fig:SQUID results}(b) shows both the experimental result and the linear fit result of the data 
which is in good agreement with theoretical linear trend shown in Fig. \ref{fig:SQUID principle}(f).
We also find that the y-intercept of linear fit result $b=-24.23(2.17)$ shifts away from 0. We attribute this difference to the
additional insertion loss along coaxial cables in dilute refrigerator and the distortion of the rising 
edge of the voltage pulse, which makes the actual amplitude of the pulse input into rf-bias line smaller than 
the set value in practice. These factors indicate that it's preferable to precompensate an additional increase in the amplitude 
so as to refer to the linear response relationship between $\Phi$ and $\Phi_e$ to modulate rf-SQUID to a specific state. 
In our system, the precompensation of the additional increase of $\Phi_e$ should be at least 24 $\Phi_0$. 
The inset figure in Fig. \ref{fig:SQUID results}(b) shows the detailed distribution of the magnetic flux transition at the 3rd data
point which is well fit by Gaussian function. 

By combining the results in Fig. \ref{fig:SQUID results}(a,b) and varying the amplitude of input pulse, we collect total 
171 local potential wells of rf-SQUID in the experiment which is shown in Fig. \ref{fig:SQUID results}(c). This result
indicates that the rf-SQUID provides qubit with maximum magnetic flux 0.32 $\Phi_0$, which tunes qubit frequency 
from maximum 5.1387 GHz to minimum 3.7721 GHz.

In addition, we compare the stability of qubit frequency obtained under the rf-SQUID scheme and the classical Z-line scheme, 
respectively. As shown in Fig. \ref{fig:SQUID results}(d), the fluctuation in unit of magnetic flux under rf-SQUID scheme with 
peak-to-peak value of $\text{P2P}=25.7$ $\mu \Phi_0$ and standard deviation of $\text{STD}=4.9$ $\mu \Phi_0$, is less than that of the 
classical Z-line scheme 
with $\text{P2P}=56.0$ $\mu \Phi_0$ and $\text{STD}=11.0$ $\mu \Phi_0$. This result indicates that the magnetic flux provided by \insitu 
rf-SQUID with RSD 24.5 PPM is more stable than the magnetic flux provided by the current in Z-line 
input from room-temperature electronics(driven by 16-bit resolution digital-to-analog converter\cite{ustc_2021_advantage}), 
whose RSD is 55.0 PPM. For the non-zero fluctuation of magnetic flux 
under the rf-SQUID scheme, the main reason is the connection between the rf-bias line and the room-temperature 
electronics resource, which transports the voltage fluctuation from room-temperature electronics to the rf-bias 
line \cite{SM}.

\section{Conclusion and Outlook}
\begin{figure}[ht]
	\includegraphics[keepaspectratio,width=8cm]{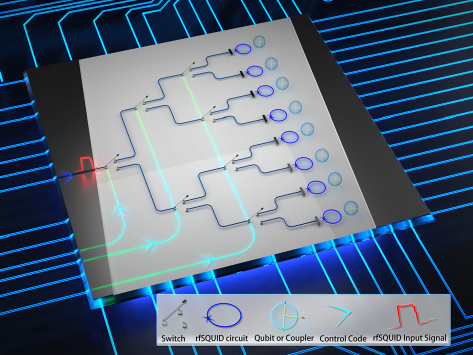}
	\caption{A time-division-multiplex (TDM) scheme of controlling qubit frequencies by combining rf-SQUID with 
	on-chip single-pole double-throw switch arrays \cite{switch_Wallraff_2016,switch_2024_PRApplied}, which 
	paves the way to scalable superconducting quantum processor. 
	With the help of switch array, the number of cables for transmitting the pulsed signal for rf-SQUID 
	could be reduced. For instance, a two-dimensional square lattice 
	quantum processor with $n$ qubits and $\sim 2n$ couplers connecting nearest neighboring qubits needs $\sim 3n$ Z-line 
	cables in usual scheme. In contrast, the number of cables could be reduced to $\sim \log_2{(3n)}+1$ with the 
	combination of rf-SQUID and switch arrays. To be specific, only one cable and one channel signal source in room temperature 
	are required in principle to transmit the pulsed signal for rf-SQUID, which could significantly reduce the power 
	consumption compared with the classical room-temperature electronics scheme, 
	and the left $\sim \log_2{(3n)}$ cables are used for transmitting the control code to address rf-SQUIDs i.e., 
	qubits or couplers. And it's straightforward to find a truth table for the control code to determine which rf-SQUID 
	to modulate.}
	\label{fig:scalable scheme with switch array}
\end{figure}
In this paper, we propose a low-power, scalable scheme to tune the transition frequency of superconducting qubit 
by using \insitu rf-SQUID. With several single square wave pulses input, the superconducting current in rf-SQUID will 
be modulated and kept at the changed position indefinitely, providing qubit with almost constant 
magnetic flux. For critical damped and overdamped condition, one single pulse input is enough to modulate rf-SQUID to
a specific state in principle. For underdamped condition, there are still some strategies to modulate rf-SQUID to a specific position
with several single pulses input. We design and fabricate a sample to investigate the properties of the underdamped 
rf-SQUID. The experimental results match well with the theoretical prediction, enabling us to modulate rf-SQUID from an 
initial state to a specific state by inputting several single square wave pulses.
In addition, we also find that the stability of qubit frequency under rf-SQUID scheme is better than that of the classical 
Z-line scheme. 
Besides the application for transmon qubit mentioned above, our scheme shall be applied for any other type of qubit whose
frequency could be modified by constant magnetic flux.
In the very near future, a shunted resistance along the JJ in the rf-SQUID is going to be fabricated to make rf-SQUID reach
overdamped condition, which will simplify the pulse inputting strategy. Moreover, by combining the rf-SQUID with switch arrays
as shown in Fig \ref{fig:scalable scheme with switch array},
the number of cables in dilute refrigerator for Z-control could be reduced from usual $\sim 3n$ to $\sim \log_2{(3n)}+1$
and only one room-temperature electronics for modulating rf-SQUID is needed in principle, which could significantly 
reduce the power consumption, thus paving the way to low-power and scalable control of large-scale superconducting processors. 

The authors thank the USTC Center for Micro- and Nanoscale Research and Fabrication for supporting the 
sample fabrication and QuantumCTek Co., Ltd. for supporting the fabrication and maintenance of 
room-temperature electronics. This research was supported by the Innovation Program for Quantum Science 
and Technology (Grant No.~2021ZD0300200), Shanghai Municipal Science and Technology Major Project 
(Grant No.~2019SHZDZX01), Anhui Initiative in Quantum Information Technologies, the New Cornerstone 
Science Foundation through the XPLORER PRIZE, the Key-Area Research and Development Program of 
Guangdong Province (Grant No.~2020B0303060001), Special funds from Jinan science and Technology 
Bureau and Jinan high tech Zone Management Committee. Y.X. acknowledges support from the Shanghai Sailing Program.

\bibliography{reference}

\pagebreak
\widetext
\begin{center}
\textbf{\large Supplemental Material for \\
``\textit{In situ} Qubit Frequency Tuning Circuit for Scalable Superconducting Quantum Computing: Scheme and Experiment"}
\end{center}
\setcounter{equation}{0}
\setcounter{figure}{0}
\setcounter{table}{0}
\setcounter{page}{1}
\makeatletter
\renewcommand{\theequation}{S\arabic{equation}}
\renewcommand{\thefigure}{S\arabic{figure}}

\section{Underdamped condition}
\label{sec_underdamped}
\begin{figure}[htbp]
	\centering
	\includegraphics[keepaspectratio,height=6cm]{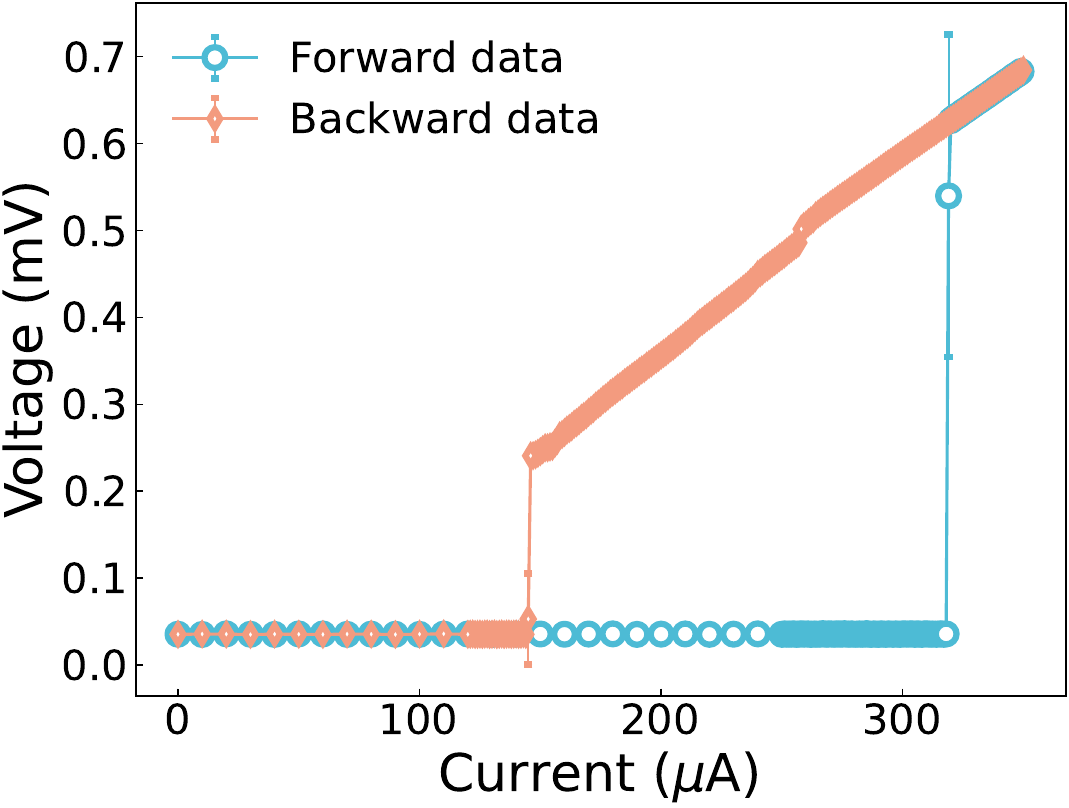}
	\caption{ The I-V curve of a JJ on the bottom chip, which is fabricated simultaneously with the JJ in rf-SQUID with
	the same fabrication parameters. The blue circles present the data obtained while increasing the current from 0 to $350$ $\mu A$, 
	while the orange diamonds present the data while decreasing the current from 350 $\mu A$ to 0.
	The Josephson critical current is $I_c = 320$ $\mu A$, the retrapping current is $I_r=146$ $\mu A$ and the slope of 
	the R branch by linear fitting is $R=1.85$ $\Omega$.}
	\label{fig:I-V data}
\end{figure}

In this section, we calculate the damping condition of the rf-SQUID. The experimental I-V data in Fig. \ref{fig:I-V data} 
shows that the Josephson critical current is $I_c = 320$ $\mu A$, the retrapping current is $I_r=146$ $\mu A$ and the slope 
by linear fitting the R branch is $R=1.85$ $\Omega$. According to these values, we could get the shunted capacitance 
$C=2.34$ pF, the parameter $\beta=1/R\sqrt{L/C}=12.1$, and the critical quality factor $\beta_c=40.1$ 
\cite{Tinkham_introduction_2014, Likharev_RMP_1979}. Since $\beta=12.1<\beta_c=40.1$, the rf-SQUID is underdamped.

\section{Thermal excitation rate}
The lifetime of the state of rf-SQUID could be affected by the thermal excitation rate. According to ref 
\cite{Voss1981}, the thermal excitation rate is
\begin{equation}
  \Gamma = \dfrac{\omega_{p,i}}{2\pi} \cdot e^{-\dfrac{H(i)}{k_B T}}
\end{equation}
where $\omega_{p,i} = \omega_{p,0}(1-i^2)^{1/4}$, $\omega_{p,0}=\sqrt{\frac{2\pi I_c}{\Phi_0 C}}$, 
$i=I/I_c$, $H(i) = \frac{I_c \Phi_0}{\pi}(\sqrt{1-i^2}-i\arccos i)-\frac{\hbar \omega_{p,i}}{2}$ and 
$I$ presents the current in rf-SQUID. Using the data from above section, we could calculate that the thermal 
excitation rate reaches a maximum value of $\Gamma_{max}=3.68\times 10^{-285}$ when rf-SQUID lies at the maximum 
position $\Phi_{max}\approx 180.8 \Phi_0$, which indicates that the lifetime of 
the metastable state of rf-SQUID is nearly infinite at the typical temperature of 20mK. Even the environmental temperature 
rises up to 100mK, the maximum thermal excitation rate reaches $\Gamma_{max}=3.82\times 10^{-49}$, which is still low enough.
The low thermal excitation rate shall be attributed to the high plasma frequency $\omega_{p,0}=2\pi \times 102.6$ GHz of rf-SQUID,
which indicates that the local potential well is quite deep.

\section{Influence of environmental magnetic flux}
\begin{figure}[htbp]
	\centering    
	\includegraphics[keepaspectratio,height=6cm]{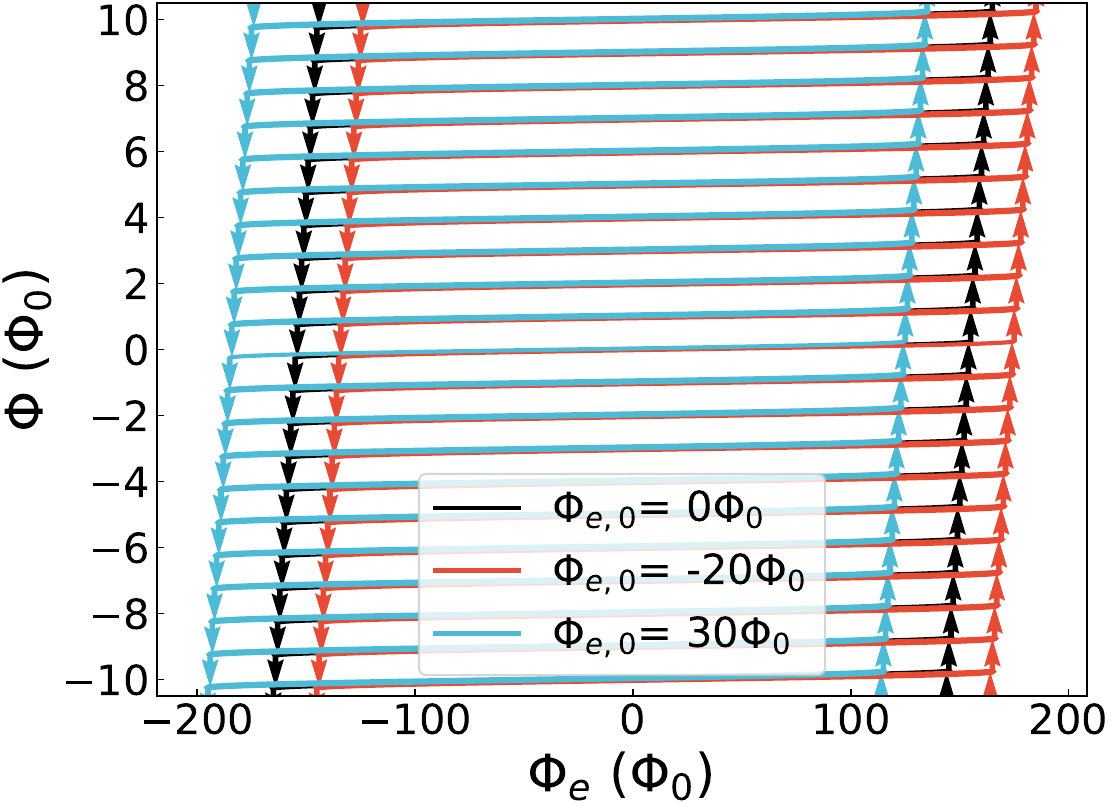}
	\caption{ Hysteresis curves affected by different constant environmental magnetic flux. Here 
	the environmental magnetic flux $\Phi_{e,0}$ are set to 0, -20, 30 $\Phi_0$ with $I_c=100$ $\mu A$, $L=1$ $nH$,
	$\beta_e=303.9$.}
  \label{fig:hysteresis curve affected by EM}
\end{figure}

The hysteresis relationship between total magnetic flux $\Phi$ of the rf-SQUID and the external magnetic flux $\Phi_e$ 
could be affected by the constant environmental magnetic flux. Fig. \ref{fig:hysteresis curve affected by EM} shows 
an example of the hysteresis curves deviated by the constant environmental magnetic flux. The existence of 
the environmental magnetic flux shall affect the magnetic flux threshold i.e., the voltage amplitude
in experiment when modulating rf-SQUID. Thanks to the truth that 
the environmental magnetic flux is often constant in practice, the influence of the environmental magnetic flux is limited.
There is around 30.5 $\Phi_0$ environmental magnetic flux in the linear magnetic flux transition experiment of the main text.

\section{Fluctuation of voltage output from room-temperature electronics}
\begin{figure}[htbp]
	\centering    
	\includegraphics[keepaspectratio,height=6cm]{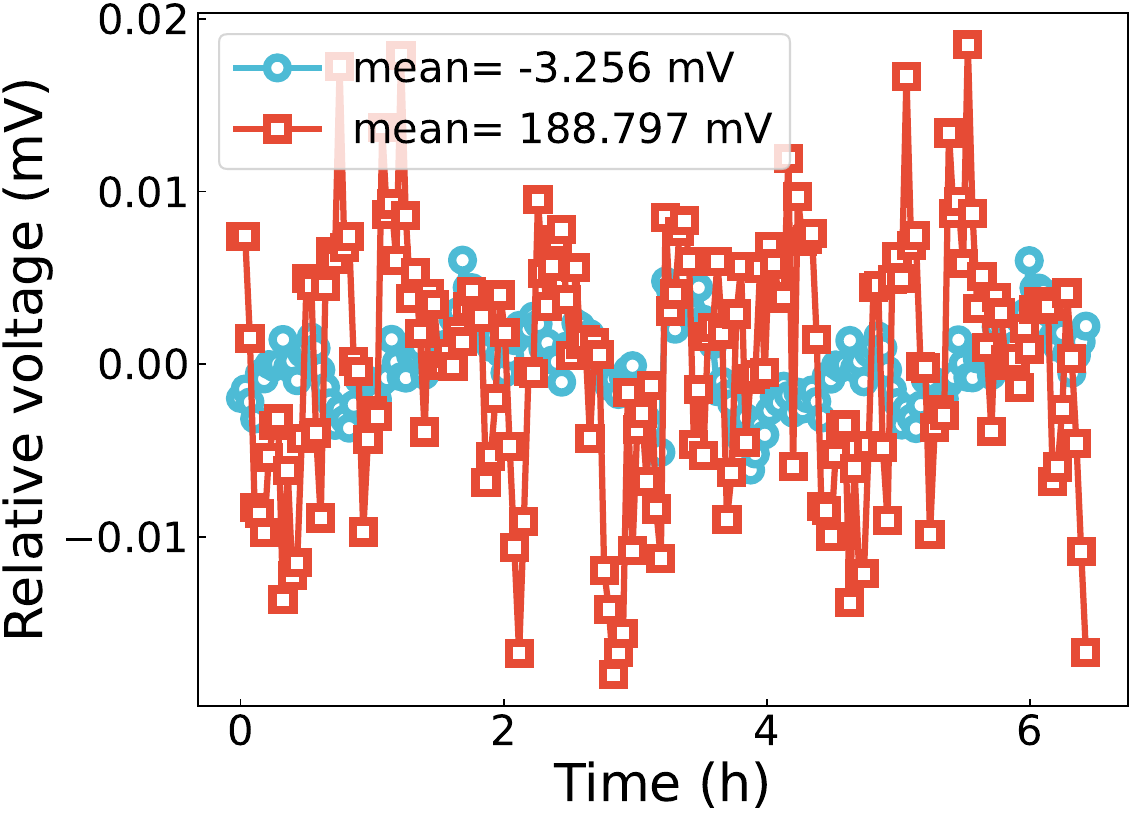}
	\caption{ A typical voltage fluctuation of room-temperature electronics resource. The vertical axis presents
	the relative voltage difference from the mean voltage. The blue circles present the data with 
	mean voltage -3.256 mV, whose peak-to-peak(P2P) value is 0.012 mV. The red squares present the data with mean voltage 
	188.797 mV, whose P2P is 0.036 mV. Each data is averaged within 1 minute and the time interval is 2 minutes.
	}
    \label{fig:voltage of room temperature electronics}
\end{figure}

For rf-SQUID scheme, there is still non-zero fluctuation of qubit frequency in the main text. The main reason is the connection
between the on-chip rf-bias line and the room-temperature electronics resource in experiment, which shall transport 
the voltage fluctuation from room-temperature electronics to the rf-bias line. A typical result of voltage fluctuation 
from room-temperature electronics is shown in Fig. \ref{fig:voltage of room temperature electronics}. When 
the voltage output approaches zero i.e., -3.256 mV, the fluctuation in voltage with $\text{P2P}=0.012$ mV will cause fluctuation 
in magnetic flux for qubit with $\text{P2P}=15.3$ $\mu\Phi_0$. When electronics resource outputs voltage 188.797 mV, 
which changes $f_{01}$ from maximum 5.0069 GHz to 4.4727 GHz in the main text, 
the fluctuation in voltage increases to $\text{P2P}=0.036$ mV which will cause the fluctuation in magnetic flux for qubit with 
$\text{P2P}=50.2$ $\mu\Phi_0$.

\end{document}